\documentclass{article}
\usepackage{ijcai24}

\usepackage{times}
\usepackage{soul}
\usepackage{url}
\usepackage[hidelinks]{hyperref}
\usepackage[utf8]{inputenc}
\usepackage[small]{caption}
\usepackage{graphicx}
\usepackage{amsmath}
\usepackage{amsthm}
\usepackage{booktabs}
\usepackage{algorithm}
\usepackage{algorithmic}
\usepackage{amsfonts}
\usepackage{multirow}
\usepackage[edges]{forest}
\usepackage{amsmath}
\usepackage[switch]{lineno}

\title{A Survey of Financial AI: Architectures, Advances and Open Challenges}

\author{
    Junhua Liu\\
    \affiliations
    Forth AI\\
    \emails
    j@forth.ai
}

\begin{document}

\maketitle

\begin{abstract}
    Financial AI empowers sophisticated approaches to financial market forecasting, portfolio optimization, and automated trading. This survey provides a systematic analysis of these developments across three primary dimensions: predictive models that capture complex market dynamics, decision-making frameworks that optimize trading and investment strategies, and knowledge augmentation systems that leverage unstructured financial information. We examine significant innovations including foundation models for financial time series, graph-based architectures for market relationship modeling, and hierarchical frameworks for portfolio optimization. Analysis reveals crucial trade-offs between model sophistication and practical constraints, particularly in high-frequency trading applications. We identify critical gaps and open challenges between theoretical advances and industrial implementation, outlining open challenges and opportunities for improving both model performance and practical applicability\footnotemark.

\footnotetext{Full list of papers and summary slides are available at: \href{https://github.com/junhua/awesome-finance-ai-papers}{https://github.com/junhua/awesome-finance-ai-papers}.}
\end{abstract}

\begin{figure*}[t]
    \centering
    \includegraphics[width=\linewidth]{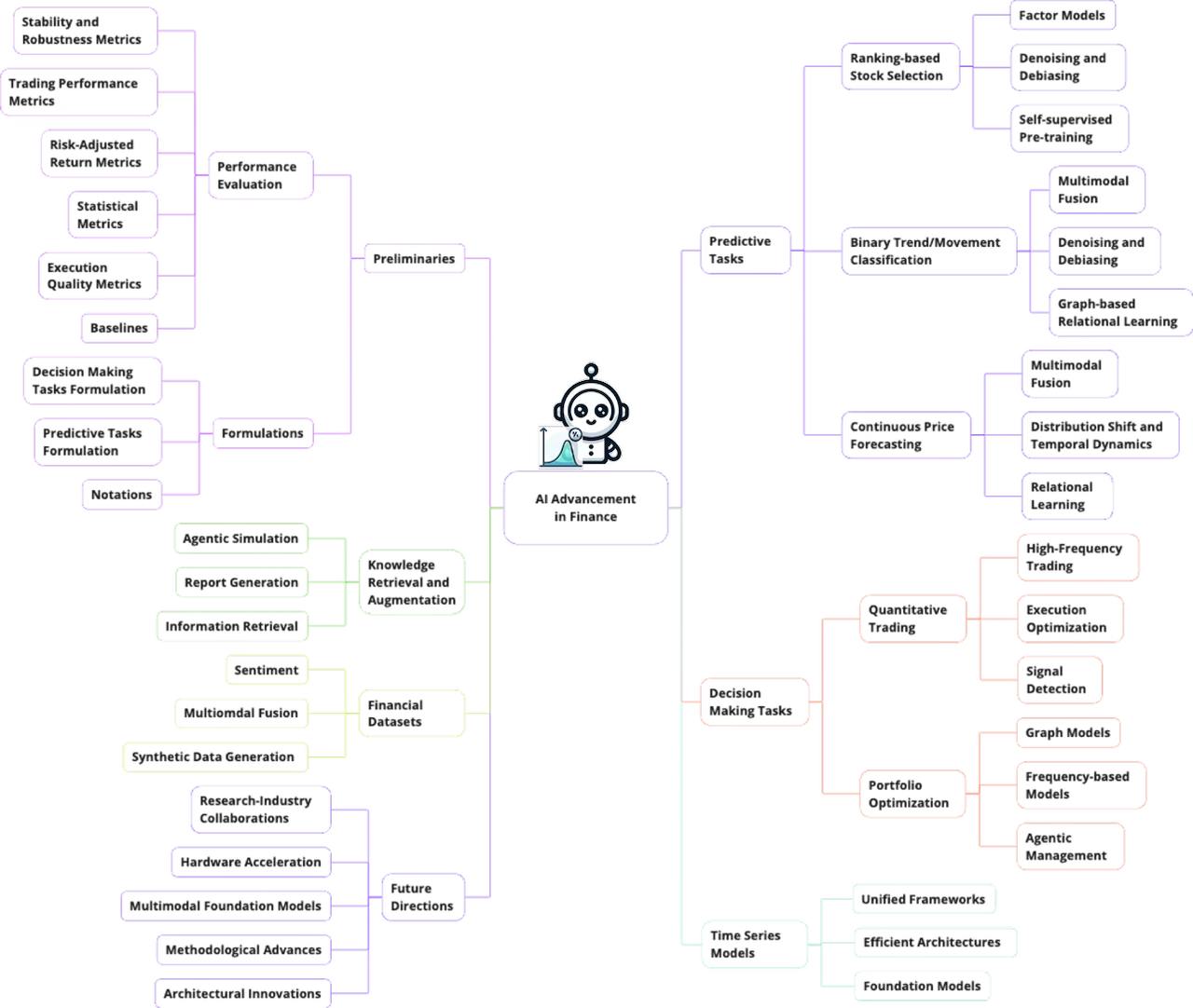}
    \caption{Taxonomy of the survey paper}
    \label{fig:taxonomy}
\end{figure*}
\section{Introduction}
Recent advancements in artificial intelligence, particularly large language models (LLMs), have significantly transformed quantitative finance. These innovations span multiple domains including predictive modeling, decision making, and knowledge retrieval, enabling more sophisticated approaches to market analysis and trading automation. While Large Language Models (LLMs) have garnered significant attention for their capabilities in natural language processing and reasoning, parallel developments in specialized architectures for financial applications demonstrate equal innovation and practical impact.

Classical approaches to financial modeling face fundamental limitations in capturing complex market dynamics, handling non-stationary distributions, and integrating diverse information sources. Recent work addresses these challenges through three primary directions: architectural innovations in deep learning models, methodological advances in training and optimization, and practical improvements in deployment and scalability. These developments enable more robust prediction under market uncertainty, more efficient portfolio optimization under constraints, and more sophisticated trading strategies incorporating multiple information sources.

\subsection{Related Work}
Several recent surveys have explored the applications of LLMs in finance from different perspectives. \cite{lee2024survey} present the first comprehensive review of financial LLMs (FinLLMs) including their evolution from general-domain LMs to financial-domain LMs, and provide extensive coverage of model architectures, training data, and benchmarks. \cite{li2023large} review the current approaches employing LLMs in finance and propose a decision framework to guide their adoption, focusing on practical implementation considerations. \cite{dong2024scoping} provide a scoping review on ChatGPT and related LLMs specifically in accounting and finance, highlighting use cases in these domains. \cite{zhao2024revolutionizing} comprehensively explore the integration of LLMs into various financial tasks and applications. \cite{ding2024large} focus specifically on LLM agents in financial trading, examining architectures, data inputs, and empirical performance through systematic analysis of 27 relevant papers. 

Figure~\ref{fig:taxonomy} shows the taxonomy of our work, which provides greater breadth across financial applications while maintaining rigorous depth in specific areas, as compared to the current surveys. Undoubtedly, these surveys have advanced our understanding of LLMs in finance. However, there remain significant gaps in addressing the broader implications for financial decision-making and industry practices. Notably, many surveys lack detailed analysis of real-world deployment challenges, integration with existing systems, and comprehensive evaluation frameworks that consider both technical and practical aspects. Our survey addresses these gaps by providing an expanded view of implementation challenges and opportunities.

\subsection{Contributions}
Our survey makes four primary contributions that distinguish it from prior work:

First, we provide a systematic analysis of the common formulations, techniques, evaluations across the complete spectrum of financial AI applications. The analysis results in a comprehensive task-oriented categorization of the surveyed works.

Second, we present a comprehensive review of architectural and methodological advances beyond LLMs, including wide variations of Graph Neural Networks, Reinforcement Learning, and time series architectures, providing insights for optimal model selection based on specific financial applications.

Third, we identify critical gaps between theories and applications, bridging the gap between academia and industry, and inspiring future research investigation for both researchers and practitioners in advancing the field.

Through rigorous examination of recent work across both academic literature and industry applications, we demonstrate how innovations in AI architecture and methodology enable more sophisticated approaches to financial modeling while highlighting crucial areas for future research. Our analysis emphasizes practical implementation considerations alongside theoretical advancements, providing a uniquely comprehensive resource for researchers and practitioners in computational finance.

\subsection{Survey Organization}
The survey is organized into eleven core sections. Section~\ref{sec:prelim} establishes the mathematical preliminaries and notations for both predictive and decision-making tasks in financial markets. Section~\ref{sec:eval} presents a comprehensive evaluation framework encompassing statistical accuracy metrics and trading performance measures. Sections~\ref{sec:price-forecast} to~\ref{sec:rank} examine predictive tasks: continuous price forecasting, binary trend classification, and ranking-based stock selection, analyzing architectural innovations and empirical validations. Section~\ref{sec:portfolio} and~\ref{sec:quant} investigate decision-making tasks spanning portfolio optimization and quantitative trading, with emphasis on multi-agent frameworks and execution strategies. Section~\ref{sec:knowledge} explores knowledge retrieval and augmentation, focusing on automated analysis and market simulation. Section~\ref{sec:data} surveys recent financial datasets and benchmarks, while Section~\ref{sec:ts} examines specialized time series models. Finally, Section~\ref{sec:challenge} identifies promising research directions and open challenges in financial AI. Throughout, we maintain rigorous analysis of theoretical foundations while emphasizing practical deployment considerations.
\section{Preliminaries}
\label{sec:prelim}

Financial market applications can be broadly categorized into predictive tasks (continuous price forecasting, binary trend classification, ranking-based selection) and decision-making tasks (portfolio optimization, quantitative trading). Despite their distinct objectives, these tasks share common mathematical foundations and feature spaces. This section presents a unified formulation framework and notations that encompass both prediction and decision-making problems.

\subsection{Notation and Problem Setup}
Let $\mathcal{S} = \{s_1, ..., s_N\}$ denote a pool of $N$ assets. For each asset $s_i$, we observe a temporal sequence of $T$ historical features $\mathbf{X}_i = \{\mathbf{x}_i^t\}_{t=1}^T \in \mathbb{R}^{T \times d}$, with $d$ being the feature dimension. The feature vector $\mathbf{x}_i^t$ encompasses three categories of market information. The first category consists of price-derived features, including open, high, low, close prices, and trading volume, which capture direct market activities. The second category comprises technical indicators such as moving averages, relative strength index (RSI), and Bollinger bands, which provide insights into market momentum and trends. The third category includes fundamental metrics like price-to-earnings ratio, market capitalization, and liquidity measures, which reflect underlying asset value and market characteristics.

The formulation incorporates several components to capture broader market dynamics. The market state $\mathbf{M}^t \in \mathbb{R}^{d_m}$ represents broader market conditions, economic indicators, and risk factors. The relational structure $\mathcal{G} = (\mathcal{V}, \mathcal{E}, \mathcal{R})$ models complex relationships between assets, where $\mathcal{V}$ represents the set of assets, $\mathcal{E}$ captures their interactions, and $\mathcal{R}$ defines different types of relationships. Textual information $\mathcal{T}_i^t$ from sources such as news and financial reports provides qualitative context. The agent's state $\mathbf{s}_t$ includes portfolio positions $\mathbf{w}_t \in \mathbb{R}^N$, capital $B_t$, and transaction costs $\mathbf{c}_t$, essential for decision-making tasks.

\begin{table}[t]
\caption{Summary of Notations in Problem Formulations}
\label{tab:notation-prob}
\begin{center}
\vspace{-.3cm}
\begin{tabular}{cl}
\toprule
\textbf{Symbol} & \textbf{Description} \\
\midrule
$\mathcal{S}$ & Asset pool \\
$\mathbf{x}_i^t$ & Feature vector for asset $i$ at time $t$ \\
$\mathbf{X}_i$ & Temporal feature sequence for asset $i$ \\
$\mathbf{M}^t$ & Market state at time $t$ \\
$\mathcal{G}$ & Asset relationship graph \\
$\mathcal{T}_i^t$ & Textual information for asset $i$ at time $t$ \\
$\mathbf{w}_t$ & Portfolio weights at time $t$ \\
$B_t$ & Available capital at time $t$ \\
$\mathbf{c}_t$ & Transaction costs at time $t$ \\
$f_\theta$ & Prediction function with parameters $\theta$ \\
$\pi_\phi$ & Policy function with parameters $\phi$ \\
$h_\psi$ & Encoder function with parameters $\psi$ \\
$p_i^t$ & Price of asset $i$ at time $t$ \\
$r_i^t$ & Return of asset $i$ at time $t$ \\
$\delta$ & Threshold for binary classification \\
\bottomrule
\end{tabular}
\end{center}
\end{table}

\subsection{Predictive Tasks Formulations}

\subsubsection{Continuous Price Forecasting}
The continuous forecasting task learns a function $f_\theta$ mapping historical observations to future values over horizon $h$:
\begin{equation}
\hat{\mathbf{y}}_i^{t+1:t+h} = f_\theta(\mathbf{x}_i^{t-w+1:t}, \mathbf{M}^t, \mathcal{G}, \mathcal{T}_i^t)
\end{equation}

The target variable $\mathbf{y}$ can take several forms: raw price levels $y_i^t = p_i^t$, returns $y_i^t = (p_i^t - p_i^{t-1})/p_i^{t-1}$, or volatility $y_i^t = \sqrt{\mathbb{E}[(r_i^t - \mu_i^t)^2]}$, where $r_i^t$ represents the log-return and $\mu_i^t$ its mean.

\subsubsection{Binary Trend Classification}
The classification task extends the general forecasting framework by discretizing price movements into directional categories. The model learns a mapping function that predicts movement direction:
\begin{equation}
\hat{y}_i^{t+h} = f_\theta(\mathbf{x}_i^{t-w+1:t}, \mathbf{M}^t, \mathcal{G}, \mathcal{T}_i^t)
\end{equation}

with target label:
\begin{equation}
y_i^{t+h} = \begin{cases}
1 & \text{if } \frac{p_i^{t+h} - p_i^t}{p_i^t} > \delta \\
0 & \text{otherwise}
\end{cases}
\end{equation}
where $\delta$ represents a threshold parameter accounting for transaction costs and market impact.

\subsubsection{Ranking-based Selection}
The ranking task focuses on learning relative ordering of assets based on their expected future performance through a scoring function:
\begin{equation}
\hat{r}_i^{t+h} = f_\theta(\mathbf{x}_i^{t-w+1:t}, \mathbf{M}^t, \mathcal{G}, \mathcal{T}_i^t)
\end{equation}

which induces ranking:
\begin{equation}
\pi_t = \text{argsort}(\{\hat{r}_i^{t+h}\}_{i=1}^N)
\end{equation}

\subsection{Decision Making Tasks Formulation}

Portfolio optimization and quantitative trading can be formulated as sequential decision-making problems under uncertainty, typically modeled through Markov Decision Processes (MDPs) or their variants.

\subsubsection{Portfolio Optimization}
The portfolio optimization task aims to determine optimal asset allocations over time. Let $\mathbf{w}_t = [w_1^t, ..., w_N^t]$ represent portfolio weights at time $t$, where $w_i^t$ denotes the proportion of capital allocated to asset $i$. The state space $s_t \in \mathcal{S}$ encompasses market features $\mathbf{X}_t$, current portfolio weights $\mathbf{w}_{t-1}$, and available capital $B_t$:

\begin{equation}
s_t = [\mathbf{X}_t, \mathbf{w}_{t-1}, B_t]
\end{equation}

The action space $a_t \in \mathcal{A}$ defines target portfolio weights:
\begin{equation}
a_t = \mathbf{w}_t, \quad \text{s.t.} \sum_{i=1}^N w_i^t = 1, \quad w_i^t \geq 0
\end{equation}

The reward function incorporates both returns and transaction costs:
\begin{equation}
r_t = \sum_{i=1}^N w_i^t r_i^t - c \sum_{i=1}^N |w_i^t - w_i^{t-1}|
\end{equation}

The objective is to learn a policy $\pi_\theta: \mathcal{S} \rightarrow \mathcal{A}$ maximizing expected cumulative returns:
\begin{equation}
\max_\theta \mathbb{E}\left[\sum_{t=1}^T \gamma^t r_t | \pi_\theta\right]
\end{equation}

\subsubsection{Quantitative Trading}
Trading can be formulated as a POMDP where the true market state is partially observable. The agent's belief state $b_t$ combines observable market features with internal estimates of latent variables:

\begin{equation}
b_t = [\mathbf{X}_t, \mathbf{h}_t, p_t, v_t]
\end{equation}

where $\mathbf{h}_t$ represents hidden state estimates, $p_t$ is position size, and $v_t$ is remaining capital.

The action space includes both trade direction and size:
\begin{equation}
a_t = (d_t, q_t), \quad d_t \in \{-1,0,1\}, \quad 0 \leq q_t \leq Q_{max}
\end{equation}

The transition function incorporates market impact:
\begin{equation}
p_{t+1} = p_t + d_t q_t
\end{equation}
\begin{equation}
v_{t+1} = v_t - d_t q_t (p_t + \alpha q_t)
\end{equation}

where $\alpha$ models price impact. The reward balances profit against risk:
\begin{equation}
r_t = d_t q_t (p_{t+1} - p_t) - c|q_t| - \lambda \text{Risk}(p_t, q_t)
\end{equation}

Both tasks can be solved through various reinforcement learning approaches, with policy gradient methods being particularly suitable due to their ability to handle continuous action spaces and complex constraints. The choice between model-based and model-free approaches depends on the trade-off between sample efficiency and computational complexity.

\subsection{Performance Evaluation}
\label{sec:eval}

\subsubsection{Definitions and Notations}

The following notations are used throughout this section:

\begin{table}[t]
\caption{Summary of Notations In Performance Evaluation}
\label{tab:notations-eval}
\centering
\vspace{-.3cm}
\begin{tabular}{clp{5.5cm}}
\toprule
\textbf{Symbol} & \textbf{Description} \\ 
\midrule
$R_p$ & Portfolio return \\
$R_f$ & Risk-free rate (3-month Treasury bill yield) \\
$R_m$ & Market return (S\&P 500 index) \\
$R_b$ & Benchmark return (relevant market index) \\
$\sigma_p$ & Portfolio volatility \\
$\sigma_{p-b}$ & Relative portfolio volatility \\
$\sigma_{\text{strategy}}$ & Strategy volatility \\
$\sigma_{\text{benchmark}}$ & Benchmark volatility \\
TP & True Positive \\
TN & True Negative \\
FP & False Positive \\
FN & False Negative \\
Precision & TP / (TP + FP) \\
Recall & TP / (TP + FN) \\
$\mathbb{E}$ & Expected value operator \\
$w_i^t$ & Weight of asset $i$ at time $t$ \\
$q_t$ & Trading quantity at time $t$ \\
$P_t^e$ & Expected execution price at time $t$ \\
$P_t^a$ & Actual execution price at time $t$ \\
\bottomrule
\end{tabular}
\end{table}

\subsubsection{Statistical Metrics}

For continuous price forecasting, prediction quality is measured through:

\begin{equation}
\text{MSE} = \frac{1}{N}\sum_{i=1}^N(\hat{y}_i - y_i)^2
\end{equation}

\begin{equation}
\text{IC} = \text{corr}(\hat{\mathbf{y}}, \mathbf{y})
\end{equation}

Binary classification employs:

\begin{equation}
\text{MCC} = \frac{\text{TP} \cdot \text{TN} - \text{FP} \cdot \text{FN}}{\sqrt{(\text{TP}+\text{FP})(\text{TP}+\text{FN})(\text{TN}+\text{FP})(\text{TN}+\text{FN})}}
\end{equation}

\begin{equation}
\text{F1} = \frac{2 \cdot \text{Precision} \cdot \text{Recall}}{\text{Precision} + \text{Recall}}
\end{equation}

\subsubsection{Risk-Adjusted Return Metrics}

Portfolio performance is assessed through:

\begin{equation}
\text{SR} = \frac{R_p - R_f}{\sigma_p}
\end{equation}

\begin{equation}
\text{CR} = \frac{R_p - R_f}{\text{MDD}}
\end{equation}

where Maximum Drawdown (MDD) captures downside risk:

\begin{equation}
\text{MDD} = \max_{t}\{\max_{\tau \leq t}(V_\tau) - V_t\}/\max_{\tau \leq t}(V_\tau)
\end{equation}

Alpha measures excess returns over a benchmark:

\begin{equation}
\alpha = R_p - [\beta(R_m - R_f) + R_f]
\end{equation}

The Information Ratio quantifies consistency of excess returns:

\begin{equation}
\text{IR} = \frac{R_p - R_b}{\sigma_{p-b}}
\end{equation}

Long-term performance is captured through:

\begin{equation}
\text{AR} = \frac{252}{T}\sum_{t=1}^T r_t
\end{equation}

\begin{equation}
\text{CAGR} = (1 + R_T)^{252/T} - 1
\end{equation}

Volatility-adjusted measures include:

\begin{equation}
\lambda_{\text{Vol}} = \frac{\sigma_{\text{strategy}}}{\sigma_{\text{benchmark}}}
\end{equation}

\subsubsection{Trading Performance Metrics}

Trading strategy evaluation incorporates:

\begin{equation}
\text{Net Return} = \sum_{t=1}^T \left(r_t - c|q_t| - \alpha q_t^2\right)
\end{equation}

The Win-Loss Ratio measures trading effectiveness:

\begin{equation}
\text{WLR} = \frac{\text{Number of Profitable Trades}}{\text{Number of Loss-Making Trades}}
\end{equation}

\subsubsection{Execution Quality Metrics}

Portfolio optimization strategies are evaluated on rebalancing efficiency:

\begin{equation}
\text{Turnover} = \frac{1}{T}\sum_{t=1}^T \sum_{i=1}^N |w_i^t - w_i^{t-1}|
\end{equation}

Trading execution quality is measured through Implementation Shortfall:

\begin{equation}
\text{IS} = \frac{1}{T}\sum_{t=1}^T (P_t^e - P_t^a)q_t
\end{equation}

where $P_t^e$ and $P_t^a$ are expected and actual execution prices.

\subsubsection{Stability and Robustness Metrics}

Strategy stability is assessed through portfolio diversification measures:

\begin{equation}
\text{Effective N} = \frac{1}{\sum_{i=1}^N w_i^2}, \quad \text{HHI} = \sum_{i=1}^N w_i^2
\end{equation}

Robustness across market regimes is evaluated through conditional metrics:

\begin{equation}
\text{Regime SR} = \frac{R_p^k - R_f}{\sigma_p^k}, \quad k \in \{\text{bull}, \text{bear}, \text{neutral}\}
\end{equation}

Strategy timing ability is assessed through the Treynor-Mazuy model:

\begin{equation}
R_p - R_f = \alpha + \beta(R_m - R_f) + \gamma(R_m - R_f)^2 + \epsilon
\end{equation}

Multi-factor performance attribution employs the Fama-French framework:

\small
\begin{equation}
R_p - R_f = \alpha + \beta_M(R_m - R_f) + \beta_S\text{SMB} + \beta_H\text{HML} + \beta_M\text{MOM} + \epsilon
\end{equation}

\subsubsection{Baselines}

Performance is benchmarked against:

\begin{itemize}
\item Traditional strategies: Equal-weight, Minimum Variance, Maximum Sharpe
\item Market indices: S\&P 500, NASDAQ Composite
\item Statistical models: ARIMA, GARCH
\item Machine learning baselines: LSTM, GRU, Transformer
\end{itemize}

All metrics are computed using walk-forward optimization with appropriate lookback windows to prevent forward-looking bias.

\begin{table*}[t]
\caption{Summary of Recent Contributions in Continuous Price Forecasting}
\label{tab:contributions}
\centering
\vspace{-.3cm}
\begin{tabular}{p{3cm}p{4.5cm}p{4cm}p{4.5cm}}
\toprule
\textbf{Work} & \textbf{Key Innovation} & \textbf{Methodology} & \textbf{Primary Results} \\
\midrule
MDGNN \newline \cite{xu2024garch} & Multi-relational graph incorporating industry and institutional relationships & Hierarchical graph embedding with ALIBI position encoding & IC improvement: 15\% (CSI300) \newline Stronger performance on larger datasets \\
\midrule
MASTER \newline \cite{li2024master} & Dynamic stock correlation modeling through alternating attention & Market-guided gating mechanism for feature selection & Ranking metrics: +13\% \newline Portfolio metrics: +47\% \newline (CSI300, CSI800) \\
\midrule
DoubleAdapt \newline \cite{zhao2023doubleadapt} & Dual adaptation for distribution shifts in streaming data & Multi-head transformation layers for feature and label adaptation & Consistent improvement across LSTM, GRU, Transformer implementations \newline Model-agnostic adaptability \\
\midrule
DPA-STIFormer \newline \cite{yan2024double} & Feature-centric temporal modeling & Double-path mechanism with adaptive gating & Superior performance across CSI500, CSI1000, NASDAQ, NYSE \newline Improved IC and Sharpe Ratio \\
\midrule
DANSMP \newline \cite{zhao2022stock} & Executive-level market knowledge graph & Dual attention network for complex entity interactions & Returns: 16.97\% \newline Sharpe Ratio: 4.628 \newline (CSI300E) \\
\midrule
GINN \newline \cite{xu2024garch} & Integration of GARCH theory with neural networks & GARCH-based regularization in LSTM loss function & Consistent outperformance across 7 global indices \newline Strong zero-shot performance \\
\midrule
DIFFSTOCK \newline \cite{daiya2024diffstock} & Diffusion models for market prediction & Adaptive noise scheduling for stock volatility & Sharpe Ratio improvement: \newline NYSE: +7.92\% \newline NASDAQ: +6.18\% \\
\midrule
\cite{wang2024news} & LLM integration for news processing & Text-based transformation of numerical forecasting & Superior performance in event-driven market shifts \newline Multi-domain applicability \\
\bottomrule
\end{tabular}
\end{table*}

\section{Continuous Price Forecasting}
\label{sec:price-forecast}
Continuous price forecasting represents a fundamental challenge in quantitative finance, combining elements of time series analysis, market microstructure, and behavioral finance. Recent works in AI have led to tremendous improvements in forecasting accuracy and robustness, particularly through the integration of complex market relationships, adaptive learning mechanisms, and multimodal data sources. As summarized in Table~\ref{tab:contributions}, these works center around three primary areas, namely, relational learning approaches that capture market structure, distribution shift modeling that addresses temporal dynamics, and hybrid approaches that integrate multiple information modalities. The contributions range from theoretical innovations in model architecture (e.g., MASTER's dynamic correlation modeling) to practical applications in performance metrics (e.g., DIFFSTOCK's improvements in Sharpe ratios). This section examines these developments systematically, beginning with a rigorous mathematical formulation of the forecasting problem, followed by detailed analysis of methodological innovations and their empirical validation across different market contexts.

\subsection{Relational Learning Approaches}
The complex interdependencies in financial markets naturally motivate graph-based and relational learning approaches, which can capture both explicit market relationships and implicit correlations between assets. Recent works have demonstrated the effectiveness of such approaches through increasingly sophisticated architectures. MDGNN~\cite{xu2024garch} established a fundamental framework by modeling multi-relational market structure through a hierarchical graph representation. The framework employs a transformer structure with ALIBI position encoding to capture temporal evolution patterns, demonstrating superior performance on China's CSI300 index with significant improvements in Information Coefficient and Cumulative Return metrics. 

Extending this relational modeling paradigm, MASTER~\cite{li2024master} introduced a more dynamic approach through a specialized transformer architecture. Unlike MDGNN's static graph structure, MASTER alternates between intra-stock and inter-stock information aggregation, enabling the capture of time-varying relationships. Its market-guided gating mechanism dynamically selects relevant features based on market conditions, achieving a 13\% improvement in ranking metrics and 47\% in portfolio-based metrics on CSI300 and CSI800 markets. 

DANSMP~\cite{zhao2022stock} further advanced the field by incorporating higher-order market relationships through a comprehensive market knowledge graph (MKG). The model expands beyond traditional asset-to-asset relationships by integrating executive entities and implicit connections, processing these diverse signals through dual attention mechanisms. This richer relational modeling led to superior performance on the CSI300E dataset, achieving investment returns of 16.97\% with Sharpe ratios of 4.628, demonstrating the value of incorporating executive-level relationships in prediction models.

\subsection{Distribution Shift and Temporal Dynamics}
The non-stationary nature of financial markets presents a fundamental challenge to traditional machine learning approaches, spurring the development of methods specifically designed to handle temporal distribution shifts. DoubleAdapt~\cite{zhao2023doubleadapt} addressed this challenge through a meta-learning framework that implements dual adaptation mechanisms. The framework's multi-head transformation layers adapt both features and labels into locally stationary distributions, while its model adapter learns initialization parameters that enable quick adaptation to new data. This approach showed consistent improvements across various time-series architectures, effectively handling both gradual and sudden market shifts.

Building on these insights, DPA-STIFormer~\cite{yan2024double} introduced a novel perspective on temporal modeling by treating features rather than time steps as tokens. Its double-path mechanism adaptively learns stock relationships, while a specialized decoder decomposes predictions into mean and deviation components. This architecture proved particularly effective across diverse market conditions, demonstrating robust performance on four major markets (CSI500, CSI1000, NASDAQ, NYSE).

DIFFSTOCK~\cite{daiya2024diffstock} approached the distribution shift challenge from a generative modeling perspective, leveraging denoising diffusion models. The framework's Masked Relational Transformer architecture processes different relationship types through separate attention heads, while its adaptive noise schedule accounts for both individual stock volatility and intra-cluster dynamics. This approach greatly improves Sharpe ratios (7.92\% and 6.18\% increases for NYSE and NASDAQ respectively), though with increased computational requirements.

\subsection{multimodal Fusion}
The integration of traditional financial theory with modern machine learning techniques has emerged as a promising direction for improving forecast robustness. GINN~\cite{xu2024garch} exemplifies this approach by bridging classical GARCH models with neural networks. The framework incorporates GARCH predictions as a regularization term in the LSTM's loss function, effectively combining statistical finance theory with deep learning flexibility. Evaluated on seven global market indices over a 30-year period, both GINN and its variant GINN-0 demonstrated consistent outperformance over traditional approaches, though with a trade-off in capturing market volatility.

Taking a different approach to multimodal integration,~\cite{wang2024news} leveraged recent works in large language models to incorporate unstructured news data into forecasting. The framework transforms numerical forecasting into a text-based task, employing LLM-based agents for news filtering and prediction evaluation. This approach proved particularly effective in capturing sudden market shifts caused by external events, demonstrating the value of incorporating qualitative information in traditionally quantitative predictions.

\begin{table*}[t]
\caption{Comparison of Recent Continuous Price Forecasting Approaches}
\label{tab:comparison-analysis}
\begin{center}
\vspace{-.3cm}
\begin{tabular}{ccccccp{4.8cm}}
\toprule
Model & Temporal & Relational & Distribution & multimodal & Compute & Market Coverage \\
      & Modeling & Learning   & Shift        & Integration & Cost    & (Datasets) \\
\midrule
MDGNN & \checkmark & \checkmark & - & - & Medium & China (CSI100, CSI300) \\
MASTER & \checkmark & \checkmark & \checkmark & - & High & China (CSI300, CSI800) \\
DoubleAdapt & \checkmark & - & \checkmark & - & Low & China (CSI300, CSI500) \\
DPA-STIFormer & \checkmark & \checkmark & \checkmark & - & High & China (CSI500, CSI1000), \newline US (NASDAQ, NYSE) \\
GINN & \checkmark & - & - & \checkmark & Low & Global (7 major market indices, 30-year period) \\
DIFFSTOCK & \checkmark & \checkmark & \checkmark & - & Very High & US (NASDAQ, NYSE, StockNet) \\
DANSMP & \checkmark & \checkmark & - & \checkmark & High & China (CSI100E, CSI300E) \\
\bottomrule
\end{tabular}
\end{center}

\end{table*}

\subsection{Comparative Analysis}

The reviewed approaches exhibit distinct characteristics in terms of their modeling capacity, computational efficiency, and practical applicability. Table~\ref{tab:comparison-analysis} summarizes the key aspects of representative models.

In terms of architectural design, models exhibit a clear trade-off between modeling capacity and computational efficiency. MASTER and DPA-STIFormer achieve superior performance through sophisticated attention mechanisms but require significant computational resources. In contrast, DoubleAdapt and GINN maintain efficiency through focused adaptation mechanisms and classical model integration, respectively.

Market coverage and generalization capabilities also vary significantly. While models like MDGNN and DANSMP demonstrate strong performance on Chinese markets, DPA-STIFormer and DIFFSTOCK show broader applicability across both US and Chinese markets. GINN stands out with its global market coverage, though focusing on major indices rather than individual stocks. This market specialization versus generalization represents a key consideration in model selection and development.

The integration of domain knowledge presents another key differentiator. GINN's incorporation of GARCH theory leads to more interpretable and theoretically grounded predictions, while purely data-driven approaches like DIFFSTOCK achieve higher performance at the cost of reduced interpretability.

\begin{table*}[t]
\caption{Summary of Recent Contributions in Binary Trend Classification}
\label{tab:binary_contributions}
\begin{center}
\vspace{-.3cm}
\begin{tabular}{p{3.1cm}p{4.3cm}p{4.25cm}p{4.4cm}}
\toprule
\textbf{Work} & \textbf{Key Innovation} & \textbf{Methodology} & \textbf{Primary Results} \\
\midrule
ECHO-GL \newline \cite{liu2024echo} & Heterogeneous graph learning from earnings calls data & Dual-mechanism approach combining spatial relations with stock dynamics & F1: +2.297\% \newline MCC: +15.629\% \newline (S\&P500, NASDAQ-100) \\
\midrule
MGDPR \newline \cite{you2024multi} & Dynamic relationship modeling through information entropy & Multi-relational diffusion process with parallel retention & Consistent outperformance across NASDAQ, NYSE, Shanghai (2016-2023) \\
\midrule
LARA \newline \cite{ijcai2024p678} & Two-stage denoising framework with locality-aware attention & Iterative refinement labeling for noise reduction & Precision: 59.1\% \newline Improved win-loss ratio in volatile markets \\
\midrule
MANA-Net \newline \cite{wang2024mana} & Dynamic market-news attention mechanism & Trainable sentiment aggregation optimized for prediction & P\&L: +1.1\% \newline Daily SR: +0.252 \\
\midrule
SH-Mix \newline \cite{jain2024saliency} & Hierarchical multimodal augmentation strategy & Modality-specific and span-based mixing techniques & Performance improvement: 3-7\% across tasks \\
\midrule
SEP \newline \cite{koa2024learning} & Self-reflective LLM framework for autonomous learning & Three-stage pipeline with PPO optimization & Superior performance in both prediction and explanation quality \\
\bottomrule
\end{tabular}
\end{center}
\end{table*}

\section{Binary Trend/Movement Classification}
\label{sec:trend}
Binary trend classification represents a critical task in quantitative finance that focuses on predicting directional price movements in financial markets. Unlike continuous price forecasting which aims to predict exact values, this task addresses the fundamental question of price movement direction, making it particularly relevant for trading strategies and risk management. Prediction accuracy are improved by market structure modeling, noise handling, and information fusion. These developments span three primary directions: graph-based relational modeling that captures complex market dependencies, denoising techniques that address market noise and distribution shifts, and multimodal approaches that integrate diverse information sources, as summarized in Table~\ref{tab:binary_contributions}.

\subsection{Graph-based Relational Learning}
The complex interdependencies in financial markets have motivated the development of sophisticated graph-based architectures that capture both explicit and implicit relationships between assets. ECHO-GL~\cite{liu2024echo} established a comprehensive framework for modeling market relationships through heterogeneous graph learning. The model introduces a dual-mechanism approach combining spatial relational modeling with stock dynamics modeling. The spatial component aggregates multimodal information across the graph structure, while the dynamics component captures post-earnings announcement effects through learnable stochastic processes. When evaluated on the S\&P 500 and NASDAQ-100 indices from 2018-2023, this architecture achieved a 2.297\% increase in F1 score and 15.629\% increase in MCC over baseline methods, with particularly strong performance during earnings seasons.
MGDPR~\cite{you2024multi} advanced the field by introducing dynamic relationship modeling through information entropy and signal energy. The framework's innovation lies in its multi-relational diffusion process, which adaptively learns and updates relationship strengths between assets. Implementation challenges, particularly in computational efficiency for large-scale markets, were addressed through sparse matrix operations and optimized graph convolutions. This approach enabled more effective capture of market dynamics, leading to consistent outperformance in next-day trend forecasting across NASDAQ, NYSE, and Shanghai markets over a seven-year test period from 2016-2023.

\subsection{Denoising and Debiasing}
Financial markets are characterized by high noise levels and non-stationary distributions, leading to the development of specialized denoising and adaptation techniques. LARA~\cite{ijcai2024p678} introduced a comprehensive framework combining locality-aware attention with iterative refinement. The framework's two-stage approach first identifies profitable trading opportunities through metric learning, then iteratively refines noisy labels to improve prediction robustness. Evaluated on high-frequency data from China's A-share market (2020-2023), cryptocurrency markets (2021-2023), and ETF markets (2019-2023), this architecture demonstrated remarkable resilience in volatile market conditions, achieving up to 59.1\% precision while maintaining computational efficiency through optimized attention mechanisms.
MANA-Net~\cite{wang2024mana} tackled the critical challenge of "aggregated sentiment homogenization" in financial news analysis. The model employs a dynamic market-news attention mechanism that weights news items based on their market relevance, integrating news aggregation and market prediction into a unified framework. Validated on an extensive dataset spanning 2003-2018 with over 2.7 million news items from major financial news sources, MANA-Net achieved a 1.1\% increase in Profit and Loss and a 0.252 increase in daily Sharpe ratio, demonstrating the value of sophisticated news aggregation in market prediction.
These binary trend classification approaches exhibit distinct characteristics in terms of their modeling capacity, computational efficiency, and practical applicability. Table~\ref{tab:binary_comparison} summarizes the key aspects of representative models.

\begin{table*}[t]
\caption{Comparison of Recent Binary Trend Classification Approaches}
\label{tab:binary_comparison}
\begin{center}
\vspace{-.3cm}
\begin{tabular}{ccccccp{5cm}}
\toprule
Model & Temporal & Graph & Distribution & Multimodal & Compute & Market Coverage \\
      & Modeling & Learning & Handling & Integration & Cost & (Datasets) \\
\midrule
ECHO-GL & \checkmark & \checkmark & - & \checkmark & High & US (S\&P500, NASDAQ-100) \\
MGDPR & \checkmark & \checkmark & \checkmark & - & Medium & Global (NASDAQ, NYSE, SSE) \\
LARA & \checkmark & - & \checkmark & - & Low & China (A-share), Crypto, ETFs \\
MANA-Net & \checkmark & - & \checkmark & \checkmark & Medium & US (S\&P500) \\
SH-Mix & \checkmark & - & - & \checkmark & Medium & US (S\&P500) \\
SEP & - & - & - & \checkmark & Very High & Global (Multiple Indices) \\
\bottomrule
\end{tabular}
\end{center}
\end{table*}

\subsection{Multimodal Fusion}
The integration of multiple data modalities presents unique challenges in feature alignment and data scarcity. SH-Mix~\cite{jain2024saliency} developed a hierarchical augmentation strategy operating at both local and global levels. The framework performs modality-specific mixing based on feature importance locally, while applying span-based mixing on fused representations globally. Built upon an attention-driven fusion architecture and evaluated on earnings call datasets from S\&P 500 companies (2019-2023), the approach achieved 3-7\% improvement over existing methods while demonstrating strong generalization capabilities across various multimodal tasks.
The emergence of large language models has enabled new approaches to feature extraction and prediction. SEP~\cite{koa2024learning} implements a three-stage pipeline combining summarization, explanation, and prediction components. The framework, tested on market data and financial news from 2020-2023, allows LLMs to autonomously learn and improve stock predictions without human expert intervention, demonstrating superior performance over both traditional deep learning and existing LLM approaches in both prediction accuracy and explanation quality.

\subsection{Comparative Analysis}

The architectural approaches demonstrate clear trade-offs between modeling sophistication and computational efficiency. Graph-based models like ECHO-GL and MGDPR achieve superior prediction accuracy through comprehensive market structure modeling but require significant computational resources. In contrast, LARA maintains efficiency through focused denoising mechanisms while sacrificing some modeling capacity.

Market coverage and generalization capabilities vary significantly across approaches. MGDPR demonstrates strong cross-market applicability spanning US, Chinese, and European markets, while models like MANA-Net and SH-Mix show specialized performance on US markets. SEP's LLM-based approach offers global coverage but with higher computational requirements and less predictable performance characteristics.

The integration of domain knowledge presents another key differentiator. ECHO-GL's incorporation of earnings call information and MANA-Net's sophisticated news processing lead to more interpretable predictions, while purely data-driven approaches like LARA achieve robust performance through statistical learning. This trade-off between interpretability and performance represents a crucial consideration in model selection.

\begin{table*}[t]
\caption{Summary of Recent Contributions in Ranking-based Stock Selection}
\label{tab:ranking_contributions}
\centering
\vspace{-.3cm}
\begin{tabular}{p{3.1cm}p{4.3cm}p{4.25cm}p{4.4cm}}
\toprule
\textbf{Work} & \textbf{Key Innovation} & \textbf{Methodological Contribution} & \textbf{Primary Results} \\ \midrule
CI-STHPAN \newline \cite{xia2024ci} & Channel-independent pre-training with dynamic hypergraph learning & Self-supervised framework with reversible normalization & IR: +21.3\%, IC: +15.7\%, \newline Outperforms SOTA on NYSE, NASDAQ \\ \midrule
ADB-TRM \newline \cite{ijcai2024p221} & Meta-learning framework for dual-level bias mitigation & Temporal-relational adversarial training with global distribution adaptation & Returns: +28.41\% \newline Risk-adjusted Returns: +9.53\% \newline (NYSE, NASDAQ, TSE) \\ \midrule
RSAP-DFM \newline \cite{ijcai2024p676} & Dual regime-shifting mechanism for factor modeling & Gradient-based posterior factors with adversarial learning & Factor Returns: +18.2\% \newline Robust macro-state adaptation \newline (A-share market) \\ \midrule
RT-GCN \newline \cite{10184655} & Pure convolutional temporal-relational modeling & Three-strategy information propagation with time-sensitive weighting & Returns: +40.4\% \newline Training Time: 13.4× faster \newline (NASDAQ, NYSE, CSI) \\ \bottomrule
\end{tabular}
\end{table*}

\begin{table*}[t]
\caption{Comparison of Recent Ranking-based Stock Selection Approaches}
\label{tab:ranking_comparison}
\begin{center}
\vspace{-.3cm}
\begin{tabular}{lccccl}
\toprule
Model & Pre-training & Factor & Distribution & Compute & Market Coverage \\
      & Integration & Modeling & Adaptation & Cost & (Datasets) \\
\midrule
CI-STHPAN & \checkmark & - & \checkmark & High & US (NYSE, NASDAQ) \\
RT-GCN & - & - & \checkmark & Low & Global (NYSE, NASDAQ, CSI) \\
ADB-TRM & - & - & \checkmark & Medium & Global (NYSE, NASDAQ, TSE) \\
RSAP-DFM & - & \checkmark & \checkmark & High & China (A-share) \\
\bottomrule
\end{tabular}
\end{center}
\end{table*}

\section{Ranking-based Stock Selection}
\label{sec:rank}
Ranking-based stock selection represents a fundamental task in quantitative finance that aims to order a universe of assets based on their expected performance. Unlike continuous price forecasting which predicts exact values or binary classification which focuses on directional movement, ranking approaches learn relative orderings that directly inform portfolio construction decisions. Ranking accuracy and robustness are optimized through innovations in self-supervised learning, bias mitigation, and integration with factor models. As summarized in Table~\ref{tab:ranking_contributions}, these works provide three primary directions: pre-training approaches that leverage unlabeled data, debiasing techniques that address various market biases, and hybrid approaches that combine deep learning with traditional finance theory.

\subsection{Self-supervised Pre-training}

The abundance of unlabeled financial data has motivated the development of sophisticated pre-training approaches that learn meaningful representations before fine-tuning for ranking tasks. CI-STHPAN~\cite{xia2024ci} established a comprehensive framework combining channel-independent processing with dynamic hypergraph learning. The model's innovation lies in constructing adaptive hypergraphs based on time series similarities using Dynamic Time Warping, moving beyond predefined relationships. To address distribution shifts, the framework incorporates reversible instance normalization and employs a two-stage training process. When evaluated on NASDAQ and NYSE markets over five years, this architecture achieved 21.3\% improvement in Information Ratio and 15.7\% in Information Coefficient, with particularly strong performance in capturing complex market dependencies.

RT-GCN~\cite{10184655} advanced the field through a pure convolutional approach to temporal-relational modeling. The framework's key innovation lies in its unified treatment of temporal patterns and stock relationships through a relation-temporal graph structure. The model employs three relation-aware strategies for information propagation - uniform, weighted, and time-sensitive - with adaptive weighting based on temporal dynamics. This architecture enables significantly faster training while maintaining high accuracy, achieving up to 40.4\% improvement in investment returns while reducing computational time by 13.4× compared to existing methods across NASDAQ, NYSE, and CSI markets.

\subsection{Denoising and Debiasing}

Financial markets exhibit various biases and non-stationary distributions that challenge traditional learning approaches. ADB-TRM~\cite{ijcai2024p221} introduced a comprehensive framework addressing both micro-level biases in stock data and macro-level distribution shifts. The model employs temporal adversarial training to handle inherent noise while using relational adversarial training to mitigate momentum spillover effects. Its meta-learning framework enables adaptation to changing market conditions through invariant feature extraction. Evaluated across NYSE, NASDAQ, and Tokyo Stock Exchange, this approach demonstrated noticeable improvements of 28.41\% in cumulative returns and 9.53\% in risk-adjusted returns while maintaining computational efficiency.

\subsection{Factor Models}

The incorporation of traditional financial theory with modern machine learning has emerged as a promising direction for improving ranking robustness. RSAP-DFM~\cite{ijcai2024p676} pioneered this approach through a dual regime-shifting mechanism that continuously captures macroeconomic states and their impact on factor dynamics. The framework employs multi-head attention for dynamic factor generation while introducing gradient-based posterior factors through adversarial learning. A novel bilevel optimization algorithm separates factor construction from model optimization, enabling more efficient training. Tested on China's A-share market, the model achieved 18.2\% improvement in factor returns while providing explicit interpretability through macroeconomic state mapping.

\subsection{Comparative Analysis}

These ranking approaches exhibit distinct characteristics in terms of their modeling capacity, computational efficiency, and practical applicability. Table~\ref{tab:ranking_comparison} summarizes key aspects of representative models.

The architectural approaches demonstrate clear trade-offs between modeling sophistication and computational efficiency. Pre-training models like CI-STHPAN achieve superior ranking accuracy through comprehensive market structure modeling but require significant computational resources. In contrast, RT-GCN maintains efficiency through focused convolutional processing while achieving competitive performance.

Market coverage and generalization capabilities vary significantly across approaches. RT-GCN and ADB-TRM demonstrate strong cross-market applicability spanning multiple global markets, while RSAP-DFM shows specialized performance on Chinese markets through its integration with factor models.

\section{Portfolio Optimization}
\label{sec:portfolio}
Portfolio optimization represents a fundamental challenge in quantitative finance that involves allocating capital across multiple assets to maximize risk-adjusted returns. Recent work in Financial AI have significantly improved portfolio management through innovations in multi-agent systems, frequency domain analysis, and risk modeling. As summarized in Table~\ref{tab:portfolio_contributions}, these works centered around three primary directions: agent-based approaches that enable adaptive management, frequency-based methods that capture market dynamics across different time scales, and network-centric techniques that model complex dependencies for risk management.

\begin{table*}[t]
\caption{Summary of Recent Contributions in Portfolio Optimization}
\label{tab:portfolio_contributions}
\begin{center}
\vspace{-.3cm}
\begin{tabular}{p{2.8cm}p{4.2cm}p{4.2cm}p{4.2cm}}
\toprule
\textbf{Work} & \textbf{Key Innovation} & \textbf{Methodological} & \textbf{Primary Results} \\ 
\midrule
MASA \newline \cite{li2024masa} & Multi-agent framework with specialized risk and return agents & Reward-based guiding mechanism for strategy diversity & Superior risk-adjusted returns across CSI300, DJIA, S\&P500 \\ 
\midrule
FreQuant \newline \cite{jeon2024frequant} & Frequency domain analysis for pattern identification & Multi-granular asset representation through DFT & ARR: 2.1× improvement, Enhanced stability in market shifts \\ 
\midrule
EarnMore \newline \cite{zhang2024reinforcement} & Customizable portfolio pools with maskable tokens & Self-supervised masking and reconstruction & Profit: +40\%, Comparable risk levels \\ 
\midrule
TrendTrader \newline \cite{ding2024trend} & Multimodal fusion of price and sentiment & Spatial-temporal RL framework & Superior ARR, ASR, MDD in DJIA, SSE-50 \\ 
\midrule
Network-EDM \newline \cite{hui2024mitigating} & Extremal risk mitigation through network theory & Maximum independent sets for diversification & Outperformance in market downturns on CSI 300 \\ 
\midrule
Market-Graph \newline \cite{yamagata2024risk} & Market graph-based clustering for index tracking & Turnover sparsity regularization with primal-dual splitting & Higher Sharpe ratios on S\&P500 \\ 
\midrule
LLM-Alpha \newline \cite{kou2024automate} & LLM-based alpha factor mining & Multi-agent system with dynamic weight-gating & Return: +53.17\% vs -11.73\% market (China, 2023) \\ 
\bottomrule
\end{tabular}
\end{center}
\end{table*}

\subsection{Comparative Analysis}

These portfolio optimization approaches exhibit distinct characteristics in terms of their modeling capacity, computational efficiency, and practical applicability. Table~\ref{tab:portfolio_comparison} summarizes key aspects of representative models.

\begin{table*}[t]
\caption{Comparison of Recent Portfolio Optimization Approaches}
\label{tab:portfolio_comparison}
\begin{center}
\vspace{-.3cm}
\begin{tabular}{lcccccl}
\toprule
\textbf{Model} & \textbf{Adaptive} & \textbf{Risk} & \textbf{Training} & \textbf{Compute} & \textbf{Market} & \textbf{Geographic/Asset} \\
&  & \textbf{Management} & \textbf{Paradigm} & \textbf{Cost} & \textbf{Coverage} & \textbf{(Datasets)} \\ 
\midrule
MASA & \checkmark & \checkmark & MARL & High & Global & CSI300, DJIA, S\&P500 \\
FreQuant & \checkmark & - & RL (DFT) & Medium & Global + Crypto & Global + Crypto \\
EarnMore & \checkmark & \checkmark & RL + SSL & Low & Multiple Markets & Multiple Markets \\
TrendTrader & \checkmark & - & RL & High & Regional & DJIA, SSE-50 \\
Network-EDM & - & \checkmark & Optimization & Medium & Regional & China (CSI 300) \\
Market-Graph & - & \checkmark & Graph Learning & Medium & Regional & US (S\&P500) \\
LLM-Alpha & \checkmark & \checkmark & LLM + RL & Very High & Regional & China (A-shares) \\
\bottomrule
\end{tabular}
\end{center}
\end{table*}

\subsection{Agentic Management}

The complexity of portfolio management has motivated the development of sophisticated multi-agent architectures that decompose the optimization problem into specialized components. MASA~\cite{li2024masa} established a comprehensive framework using three cooperative agents: a trend observer for market monitoring, a return optimizer for portfolio maximization, and a risk manager for risk minimization. The framework's key innovation lies in its reward-based guiding mechanism that combines return and action rewards to maintain strategy diversity while adapting to market conditions. When evaluated on major indices (CSI300, DJIA, S\&P500), this architecture demonstrated superior performance in both stable and volatile markets.

EarnMore~\cite{zhang2024reinforcement} advanced the field through a customizable approach to portfolio management. The framework introduces a maskable token system to represent unfavorable stocks and employs self-supervised learning for relationship modeling. Its one-shot training capability enables efficient adaptation to different stock pools while maintaining performance, achieving 40\% profit improvement across diverse market conditions.

\subsection{Frequency-based Models}

The multi-scale nature of market patterns has inspired approaches that explicitly model different frequency components. FreQuant~\cite{jeon2024frequant} pioneered this direction through a reinforcement learning framework operating in the frequency domain. The model employs Discrete Fourier Transform to identify market patterns at different frequencies, processing these through a Frequency State Encoder for multi-granular asset representation. This approach achieved up to 2.1× improvement in Annualized Rate of Return across U.S., Korean, and cryptocurrency markets, with particular strength in handling market regime shifts.

TrendTrader~\cite{ding2024trend} complemented frequency analysis with multimodal integration, combining price patterns with news sentiment through a spatial-temporal backbone network. Its incremental learning approach for portfolio weights demonstrated robust performance across DJIA and SSE-50 indices, particularly in capturing sentiment-driven market movements.

\subsection{Graph Models}

The complex dependencies in financial markets have motivated network-based approaches to risk management. \cite{hui2024mitigating} developed a framework using Graph Theory and Extreme Value Theory to mitigate extremal risks. The approach constructs graphs based on extremal dependencies between stock returns, using maximum independent sets for diversification. Tested on CSI 300 components, this method outperformed traditional sector-based approaches, especially during market downturns.

\cite{yamagata2024risk} advanced this direction by replacing sector-based diversification with market graph-based clustering. The framework incorporates turnover sparsity regularization to manage transaction costs while ensuring cluster-based diversification. Experiments on S\&P500 demonstrated superior Sharpe ratios compared to traditional methods, particularly in challenging market conditions.

The architectural approaches demonstrate clear trade-offs between modeling sophistication and computational efficiency. Multi-agent systems like MASA achieve superior performance through comprehensive market modeling but require significant computational resources. In contrast, EarnMore maintains efficiency through focused token-based processing while achieving competitive performance.

\begin{table*}[t]
\caption{Summary of Recent Contributions in Quantitative Trading}
\label{tab:quant_contributions}
\begin{center}
\vspace{-.3cm}
\begin{tabular}{p{2.5cm}p{4.5cm}p{4.5cm}p{4.5cm}}
\toprule
\textbf{Work} & \textbf{Key Innovation} & \textbf{Methodology} & \textbf{Primary Results} \\ \midrule
StockFormer & Predictive coding with multi-aspect modeling & Three-branch transformer for market dynamics & Returns: +40.3\%, SR: +22.7\% vs SAC \\ \midrule
MacMic & Order execution decomposition & Hierarchical RL with stacking HMM & Superior execution across US/CN markets \\ \midrule
IMM & Multi-price market making strategy & Adversarial state representation learning & Lower adverse selection in futures \\ \midrule
EarnHFT & Adaptive crypto trading system & Hierarchical Q-learning with router & +30\% profit vs SOTA in crypto \\ \midrule
MacroHFT & Market regime-based decomposition & Memory-enhanced policy integration & Superior risk-adjusted crypto returns \\ \midrule
DRPO & Direct portfolio weight optimization & Probabilistic state decomposition & 415ms latency in production \\ \midrule
CPPI-MADDPG & Portfolio insurance integration & Multi-agent RL with protection strategies & AR: 9.68\%, SR: 2.18 in SZSE \\ \midrule
HRT & Integrated selection-execution & PPO-DDPG hierarchy for trading & Improved diversification in S\&P500 \\ \midrule
TRR & LLM-based crash detection & Four-phase temporal reasoning & Superior detection of market crashes \\ \bottomrule
\end{tabular}
\end{center}
\end{table*}

\begin{table*}[t]
\caption{Comparison of Quantitative Trading Approaches}
\label{tab:trading_comparison}
\begin{center}
\vspace{-.3cm}
\scalebox{.85}{
\begin{tabular}{lccccp{4.2cm}p{4cm}}
\toprule
\textbf{Model} & \textbf{Strategy} & \textbf{Execution} & \textbf{Market} & \textbf{Compute} & \textbf{Key Strengths} & \textbf{Key Limitations} \\ 
\midrule
StockFormer & Predictive & Indirect & Multi-asset & High & Robust pattern extraction; \newline Volatile market performance & High compute overhead; \newline Complex training \\ 
\midrule
MacMic & Execution & Direct & Equities & Medium & Superior price execution; \newline Hierarchical control & Limited signal generation; \newline Single market focus \\ 
\midrule
IMM & Market Making & Direct & Futures & Medium & Low adverse selection; \newline Queue priority preservation & Market-specific design; \newline Limited asset coverage \\ 
\midrule
EarnHFT & HFT & Direct & Crypto & High & Strong profitability; \newline Adaptive routing & Crypto-specific; \newline Resource intensive \\ 
\midrule
MacroHFT & HFT & Direct & Crypto & High & Regime adaptation; \newline Memory-enhanced decisions & Complex training process; Market-specific \\ 
\midrule
DRPO & HFT & Direct & Multi-asset & Low & Production-ready latency; \newline Efficient computation & Limited strategy scope; \newline Basic signals \\ 
\midrule
CPPI-MADDPG & Portfolio & Indirect & Equities & Medium & Downside protection; \newline Strategy integration & Coordination overhead; Slower adaptation \\ 
\midrule
HRT & Hybrid & Direct & Equities & High & Balanced execution; \newline Portfolio diversification & Complex training; \newline High resource usage \\ 
\midrule
TRR & Signal & Indirect & Multi-asset & Very High & Novel signal sources; \newline Crisis detection & High latency; \newline Resource intensive \\ 
\bottomrule
\end{tabular}}
\end{center}
\end{table*}

\section{Quantitative Trading}
\label{sec:quant}
Quantitative trading represents a systematic approach to financial markets that leverages mathematical models and computational methods to develop and execute trading strategies. Recent works in Financial AI have significantly enhanced quantitative trading through innovations in three key areas: strategy development incorporating predictive signals and market dynamics, execution optimization through hierarchical control, and high-frequency trading systems that adapt to market microstructure. Unlike traditional approaches that rely on fixed rules or simple statistical arbitrage, modern quantitative trading systems employ sophisticated deep learning architectures to capture complex market patterns while maintaining computational efficiency for real-time deployment. As summarized in Table~\ref{tab:quant_contributions}, these works propose predictive modeling frameworks that extract meaningful signals from noisy market data, execution optimization methods that decompose complex trading decisions into manageable components, and adaptive systems that handle the unique challenges of high-frequency market environments. This section examines these developments systematically, analyzing their theoretical foundations, architectural innovations, and empirical validation across different market contexts.

\subsection{Signal Detection}

Signal Detection primarily focuses on extracting robust signals from complex market data. StockFormer~\cite{gao2023stockformer} pioneered a three-branch transformer architecture that captures long-term trends, short-term movements, and inter-asset relationships through specialized attention mechanisms. Its diversified multi-head attention design maintains pattern diversity across concurrent time series, demonstrating particular effectiveness in volatile cryptocurrency markets with 40.3\% improvement in portfolio returns.

TRR~\cite{zhang2024multimodal} represents a novel direction in incorporating qualitative data through LLMs, enabling zero-shot reasoning for unprecedented market events. Its four-phase framework combines impact graph generation, temporal context management, and relational reasoning, showing superior performance in detecting market crashes across multiple historical crisis periods.

\subsection{Execution Optimization}

Execution optimization has evolved toward hierarchical frameworks that decompose complex trading decisions. MacMic~\cite{ijcai2024p664} introduces a two-level architecture where a high-level agent handles volume scheduling while a low-level agent manages precise order placement. The framework's stacking Hidden Markov Model enables unsupervised extraction of multi-granular market representations, achieving superior price execution across US and Chinese markets.

IMM~\cite{ijcai2024p663} automates market making through multi-price level strategies, employing temporal-spatial attention networks to mitigate adverse selection risk. Its imitation learning approach from expert strategies facilitates efficient exploration of complex action spaces, demonstrating improved risk-adjusted returns in futures markets.

HRT~\cite{zhao2024hierarchical} tackles the joint problem of stock selection and execution through a hierarchical PPO-DDPG framework. Its phased alternating training ensures coordinated learning between controllers, achieving superior performance across both bullish and bearish markets while maintaining portfolio diversification.

\subsection{High-Frequency Trading}

High-frequency trading systems have advanced through specialized architectures addressing market microstructure and computational efficiency. EarnHFT~\cite{qin2024earnhft} introduces a three-stage framework combining Q-learning with specialized agent pools, effectively handling extended trading trajectories while maintaining 30\% higher profitability across market conditions.

MacroHFT~\cite{zong2024macrohft} employs a two-phase approach with specialized sub-agents for different market regimes, using a memory-enhanced hyper-agent for rapid adaptation. DRPO~\cite{han2023efficient} achieves practical deployment success through state space decomposition and probabilistic dynamic programming, maintaining 415ms latency in production environments.

CPPI-MADDPG~\cite{zhang2024optimizing} integrates classical portfolio insurance with multi-agent reinforcement learning, achieving 9.68\% annual returns while maintaining downside protection through adaptive strategy selection.

\subsection{Comparative Analysis}

The surveyed approaches demonstrate distinct characteristics in their architectural design, computational requirements, and market applicability. Table~\ref{tab:trading_comparison} summarizes key aspects across different frameworks:

The architectural approaches demonstrate clear trade-offs between modeling sophistication and practical deployment constraints. Predictive frameworks like StockFormer achieve superior performance through comprehensive market modeling but require significant computational resources. In contrast, execution-focused systems like DRPO maintain efficiency through focused optimization while sacrificing broader strategy integration.

Furthermore, compute requirements vary significantly, from DRPO's production-ready 415ms latency to TRR's resource-intensive LLM processing. This latency-sophistication trade-off critically influences deployment scenarios, with HFT systems prioritizing speed while signal generation approaches emphasize modeling capacity.

Market coverage and generalization capabilities also differ substantially. Models like StockFormer and TRR demonstrate cross-market applicability, while specialized frameworks like IMM and EarnHFT show superior performance within specific market contexts. This specialization-generalization trade-off suggests that optimal deployment strategies might involve ensemble approaches tailored to specific market conditions and trading objectives.

\section{Knowledge Retrieval and Augmentation}
\label{sec:knowledge}
Knowledge retrieval and augmentation in financial markets involves extracting, processing, and leveraging structured and unstructured information for improved decision-making. This section focuses on discussing three primary directions: financial information retrieval pipelines that processes complex financial texts, intelligent report generation that synthesizes multiple data sources, and agent-based market simulation that enables scenario analysis. As summarized in Table~\ref{tab:knowledge_contributions}, these works range from specialized architectures for financial text processing to comprehensive frameworks for market simulation.

\begin{table*}[t]
\caption{Summary of Recent Contributions in Knowledge Retrieval and Augmentation}
\label{tab:knowledge_contributions}
\centering
\vspace{-.3cm}
\begin{tabular}{p{2.8cm}p{4.2cm}p{4.2cm}p{4.2cm}}
\midrule
\textbf{Work} & \textbf{Key Innovation} & \textbf{Methodology} & \textbf{Primary Results} \\
\midrule
MACK & Matrix chunking for financial event extraction & Two-dimensional annotation method & F1: 81.33\% (Event), 96.89\% (Word) \\
\midrule
FinReport & Automated investment analysis & Three-module system with news factorization & Return: 57.76\%, Accuracy: 75.40\% \\
\midrule
LLM-Annotator & LLM-based financial relation extraction & Reliability index for annotation quality & 29\% improvement over crowdworkers \\
\midrule
TRR & Temporal reasoning for crash detection & Four-phase news analysis framework & Superior crisis detection across multiple periods \\
\midrule
StockAgent & Multi-LLM trading simulation & Dynamic agent behavior modeling & Quantified impact of information sources \\
\midrule
EconAgent & LLM-powered economic simulation & Perception-memory modules for agents & Reproduction of key economic phenomena \\
\bottomrule
\end{tabular}
\end{table*}

\subsection{Information Retrieval}

Information retrieval (IR) focused on processing complex financial documents through specialized models. MACK~\cite{huang2024extracting} established a comprehensive framework for Chinese financial event extraction through matrix chunking. Unlike previous approaches requiring pre-identified entities, MACK processes raw text directly through a two-dimensional annotation method that visualizes component interactions. When evaluated on the FINEED dataset with 5,000 annotated events, this architecture achieved 81.33\% F1-score in event extraction while maintaining 96.89\% accuracy in word segmentation.

The emergence of large language models has enabled new approaches to financial annotation. LLM-Annotator~\cite{aguda2024large} demonstrated that models like GPT-4 and PaLM 2 can outperform crowdworkers by 29\% in accuracy while maintaining cost efficiency. Using the REFinD dataset with 28,676 SEC filing relations, the framework's reliability index successfully identified instances requiring expert review, enabling reliable automation of approximately 65\% of annotation tasks.

\subsection{Report Generation}

The synthesis of multiple information sources for automated analysis has advanced through sophisticated architectures. FinReport~\cite{li2024finreport} pioneered this direction through a three-module system combining news factorization, return forecasting, and risk assessment. The framework integrates semantic role labeling with dependency parsing for news understanding, while employing enhanced Fama-French models for return prediction. This approach achieved 75.40\% accuracy in news factorization and 57.76\% annualized returns in backtesting.

TRR~\cite{koa2024temporal} advanced the field through temporal relational reasoning for market event detection. The framework's four-phase approach combines impact chain generation, temporal context management, and PageRank-based attention mechanisms. This architecture demonstrated particular effectiveness in detecting market crashes across multiple historical crises, including the 2007 financial crisis and 2020 COVID-19 crash.

\subsection{Agentic Simulation}

The complexity of market interactions has motivated the development of sophisticated simulation frameworks. StockAgent~\cite{zhang2024ai} established a comprehensive multi-agent system using different LLMs to simulate realistic trading behaviors. The framework incorporates external factors including macroeconomic conditions and market sentiment, revealing distinct trading patterns between models like GPT-3.5 and Gemini Pro. Experimental results quantified the impact of different information sources, with interest rate information significantly affecting trading frequency.

EconAgent~\cite{li2024econagent} advanced this direction through LLM-powered agents for macroeconomic simulation. The framework's innovation lies in its perception and memory modules that enable heterogeneous decision-making while considering historical market dynamics. This approach successfully reproduced key economic phenomena like the Phillips Curve and demonstrated effectiveness in simulating crisis impacts, particularly during the COVID-19 period.

\subsection{Comparative Analysis}

Table~\ref{tab:knowledge_comparison} summarizes key aspects of representative models The three research directions in knowledge retrieval and augmentation offer complementary approaches to market understanding, each with distinct theoretical limitations. Information extraction methods face fundamental challenges in handling context-dependent financial terminology and evolving market jargon. While MACK demonstrates strong performance on Chinese texts, the generalization to multiple languages and domains remains challenging due to linguistic complexity and domain-specific variations.

Report generation frameworks encounter theoretical limitations in causal reasoning and temporal dependency modeling. Although FinReport successfully combines multiple information sources, establishing robust causal relationships between news events and market movements remains an open challenge. The inherent delay between information release and market impact further complicates temporal modeling.

Simulation approaches face fundamental limitations in agent behavior modeling and market complexity. While StockAgent and EconAgent demonstrate impressive capabilities, they necessarily simplify market microstructure and agent interactions. The challenge of calibrating agent behaviors to realistic market conditions while maintaining computational tractability represents a key area for future research.

\begin{table*}[t]
\caption{Summary of Recent Contributions in Financial Datasets}
\label{tab:dataset_contributions}
\centering
\vspace{-.3cm}
\begin{tabular}{p{2.2cm}p{4cm}p{5.2cm}p{4.2cm}}
\toprule
\textbf{Dataset} & \textbf{Key Innovation} & \textbf{Data Characteristics} & \textbf{Primary Results} \\
\midrule
Market-GAN & Controllable synthetic data with semantic context & Two-stage GAN with C-TimesBlock for temporal consistency & Superior fidelity in DJIA simulation (2000-2023) \\
\midrule
FNSPID & Large-scale price-news integration framework & 29.7M prices, 15.7M news records for 4,775 companies & Transformer accuracy R² = 0.988, Reproducible updates \\
\midrule
AlphaFin & Chain-of-thought financial reasoning & Market data with expert reasoning annotations & 30.8\% annualized returns, Enhanced interpretability \\
\midrule
StockEmotions & Fine-grained investor psychology & 10k annotated comments, 12 emotion classes, emoji features & Improved prediction with emotional features \\
\bottomrule
\end{tabular}
\end{table*}

\section{Financial Datasets}
\label{sec:data}
Financial datasets play a crucial role in advancing artificial intelligence applications for market analysis and trading. Recent developments in dataset creation have focused on three primary directions: synthetic data generation for enhanced model training, multimodal integration combining market data with textual information, and specialized datasets for reasoning and sentiment analysis. Unlike traditional financial datasets that focus solely on price and volume information, modern datasets incorporate diverse data types including news sentiment, investor emotions, and chain-of-thought reasoning annotations. As summarized in Table~\ref{tab:dataset_contributions}, these works contribute to both the creation of novel datasets and the development of sophisticated data generation techniques, addressing critical challenges in data availability, quality, and comprehensiveness for financial applications.

\subsection{Synthetic Data Generation}
Market-GAN~\cite{xia2024market} established a framework for generating high-fidelity financial data with controllable semantic context through a novel two-stage training approach. The framework combines GAN architecture with autoencoder and supervisory networks to maintain data fidelity while ensuring alignment with given market contexts. Its C-TimesBlock innovation effectively captures temporal dependencies while preventing mode collapse, a common challenge in financial time series generation. When evaluated on Dow Jones Industrial Average data spanning 2000-2023, this approach demonstrated superior performance in context alignment and downstream task fidelity compared to existing generation methods.
\subsection{Multimodal Fusion}
FNSPID~\cite{dong2024fnspid} pioneered large-scale integration of quantitative and qualitative financial data through a comprehensive dataset combining 29.7 million stock prices with 15.7 million aligned news records. The dataset covers 4,775 S\&P500 companies from 1999-2023, demonstrating that increased dataset scale significantly improves prediction accuracy. Experiments with various deep learning architectures revealed that transformer-based models achieve optimal performance (R² = 0.988) when leveraging the dataset's full scale, while sentiment information provides modest but consistent improvements in prediction accuracy.
\subsection{Sentiment and Emotion}
AlphaFin~\cite{li2024alphafin} addressed the interpretability gap in financial analysis through a dataset combining traditional market data with chain-of-thought annotations. The framework's integration with retrieval-augmented generation enables real-time market awareness while maintaining reasoning capabilities, achieving 30.8\% annualized returns in experimental validation. StockEmotions~\cite{lee2023stockemotions} complemented this direction through fine-grained emotion analysis, providing 10,000 annotated StockTwits comments across 12 emotion classes. The dataset's multi-step annotation pipeline, combining pre-trained language models with expert validation, demonstrated that incorporating emotional features significantly improves market prediction accuracy compared to purely numerical approaches.

\begin{table*}[t]
\caption{Comparison of Financial Dataset Characteristics}
\label{tab:dataset_comparison}
\begin{center}
\vspace{-.3cm}
\begin{tabular}{lccccl}
\toprule
\textbf{Dataset} & \textbf{Time} & \textbf{Text} & \textbf{Market} & \textbf{Scale} & \textbf{Primary} \\
      & \textbf{Series} & \textbf{Data} & \textbf{Coverage} & & \textbf{Application} \\
\midrule
Market-GAN & \checkmark & - & DJIA & Medium & Market Simulation \\
FNSPID & \checkmark & \checkmark & S\&P500 & Very Large & Predictive Modeling \\
AlphaFin & \checkmark & \checkmark & Multiple & Large & Interpretable Analysis \\
StockEmotions & \checkmark & \checkmark & StockTwits & Medium & Sentiment Analysis \\
\bottomrule
\end{tabular}
\end{center}
\end{table*}

\subsection{Comparative Analysis}

These financial datasets exhibit distinct characteristics in their coverage, scale, and applicability to different financial tasks. As shown in Table~\ref{tab:dataset_comparison}, coverage ranges from focused sentiment analysis (StockEmotions) to comprehensive market-wide data integration (FNSPID). Scale varies significantly, with FNSPID providing millions of aligned price-news records while StockEmotions offers deeper but narrower emotional annotation.

Market coverage and temporal resolution reveal complementary strengths across datasets. Market-GAN focuses on high-fidelity synthetic data for a single major index (DJIA) with emphasis on temporal consistency. In contrast, FNSPID provides broader market coverage across the S\&P500 with emphasis on news-price alignment. AlphaFin bridges multiple markets while prioritizing reasoning annotation quality, and StockEmotions offers specialized coverage of retail investor sentiment through StockTwits data.

\section{Time Series Models}
\label{sec:ts}

\begin{table*}[t]
\caption{Summary of Recent Contributions in Time Series Models}
\label{tab:timeseries_contributions}
\centering
\vspace{-.3cm}
\begin{tabular}{p{2.8cm}p{4.2cm}p{4.2cm}p{4.2cm}}
\hline\toprule 
\textbf{Model} & \textbf{Key Innovation} & \textbf{Methodology} & \textbf{Primary Results} \\
\midrule
Timer & Pre-trained decoder-only transformer & Unified sequence format for heterogeneous data & Strong few-shot performance with 1-5\% data \\
\midrule
MOMENT & Open-source foundation model family & Multi-dataset training strategies & Superior limited-supervision performance \\
\midrule
TimeMixer & Multiscale MLP architecture & Decomposable mixing blocks for temporal patterns & SOTA across 18 benchmarks \\
\midrule
TimesNet & Multi-periodicity analysis & 1D to 2D transformation for temporal patterns & Unified performance across 5 major tasks \\
\midrule
PatchTST & Patch-based time series processing & Channel-independent transformer & 21\% MSE reduction in long-term forecasting \\
\bottomrule
\end{tabular}
\end{table*}

Time series modeling for financial applications focuses on three primary directions: foundation models that leverage large-scale pre-training, efficient architectures that handle temporal complexity, and unified frameworks that address multiple time series tasks. Unlike traditional statistical approaches, modern time series models incorporate transformer architectures, multiscale analysis, and transfer learning capabilities. As summarized in Table~\ref{tab:timeseries_contributions}, these foundation models are trained on massive datasets, specialized architectures for temporal modeling, and unified frameworks supporting multiple tasks.

\subsection{Foundation Models}
The emergence of foundation models has transformed time series analysis through large-scale pre-training and transfer learning. Timer established a comprehensive framework using a decoder-only transformer architecture trained on 1 billion time points. The model's innovation lies in its unified sequence format that handles heterogeneous time series data while enabling few-shot learning capabilities. When evaluated across forecasting, imputation, and anomaly detection tasks, this architecture achieved state-of-the-art performance using only 1-5
MOMENT advanced this direction through an open-source family of foundation models for general-purpose time series analysis. The framework addresses key challenges in multi-dataset training through specialized techniques for handling varying sampling rates and amplitudes. Its contribution of Time Series Pile, a diverse collection of public datasets, enables robust pre-training while maintaining reproducibility. This approach demonstrated particular strength in limited-supervision settings across financial applications.

\begin{table*}[t]
\caption{Comparison of Time Series Model Characteristics}
\label{tab:timeseries_comparison}
\begin{center}
\vspace{-0.3cm}
\begin{tabular}{lccccp{4.2cm}}
\toprule
\textbf{Model} & \textbf{Pre-training} & \textbf{Multi-task} & \textbf{Compute} & \textbf{Scalability} & \textbf{Primary Application} \\ 
\midrule
Timer      & \checkmark & \checkmark & High        & Linear      & Few-shot Learning \\
MOMENT     & \checkmark & \checkmark & High        & Linear      & Limited Supervision \\
TimeMixer  &    -       &     -      & Medium      & Sublinear   & Multiscale Patterns \\
TimesNet   &     -      & \checkmark & Medium      & Linear      & Task Flexibility \\
PatchTST   &     -      &     -      & Low         & Sublinear   & Long Sequences \\ 
\bottomrule
\end{tabular}
\end{center}
\end{table*}

\subsection{Efficient Architectures}
The complexity of temporal patterns has motivated development of specialized architectures balancing modeling capacity with computational efficiency. TimeMixer pioneered this direction through a multiscale MLP-based architecture that decomposes temporal variations at different sampling scales. The model employs Past-Decomposable-Mixing blocks for separate processing of seasonal and trend components, while Future-Multipredictor-Mixing blocks combine predictions across scales. This architecture demonstrated state-of-the-art performance across 18 benchmarks while maintaining computational efficiency.
PatchTST advanced efficient processing through a patch-based approach to time series modeling. The framework's key innovations lie in subseries-level patching for local information capture and channel-independent processing for univariate sequences. This architecture achieved 21\% reduction in Mean Squared Error for long-term forecasting while effectively handling extended historical sequences without memory constraints.

\subsection{Unified Frameworks}
The diverse requirements of time series analysis have inspired development of unified frameworks supporting multiple tasks. TimesNet established a comprehensive architecture through multi-periodicity analysis that transforms one-dimensional sequences into two-dimensional representations. The framework captures both intraperiod and interperiod variations through parameter-efficient inception blocks, demonstrating superior performance across forecasting, imputation, classification, and anomaly detection tasks.

\subsection{Discussion}
The three research directions in time series modeling offer complementary approaches to temporal analysis, each with distinct theoretical limitations. Table~\ref{tab:timeseries_comparison} summarizes key aspects across different frameworks. Foundation models face fundamental challenges in capturing rare temporal patterns and handling non-stationary distributions, though they excel at extracting general temporal dependencies. While large-scale pre-training improves robustness, the inherent challenge of temporal causality and regime changes remains.

Efficient architectures encounter trade-offs between model capacity and computational complexity. While patch-based approaches and multiscale decomposition provide practical solutions, they necessarily introduce approximations in temporal modeling. The balance between local and global temporal dependencies remains a key theoretical challenge.

Unified frameworks face fundamental limitations in optimal parameter sharing across diverse tasks. While architectures like TimesNet demonstrate impressive multi-task capability, the theoretical foundations for optimal architecture design across different temporal modeling objectives remain incomplete. These limitations suggest future research directions in developing more theoretically grounded approaches to temporal modeling while maintaining practical applicability. 

The integration of domain knowledge remains crucial for developing more robust and interpretable temporal models, particularly in financial applications where model reliability and theoretical soundness are paramount.

\section{Open Challenges}
\label{sec:challenge}

Having comprehensively reviewed the recent works on Financial Ai, we identify several promising research directions that warrant further investigation. In this section, We organize these opportunities into architectural innovations, methodological advancements, and practical considerations to potentially extend the impact to further advancement of the field.

\subsection{Architectural Innovations}
The evolution of financial AI architectures suggests several key directions for improvement. Foundation models pre-trained on massive financial datasets show promise~\cite{liu2024timer,goswami2024moment} but currently lack domain-specific inductive biases crucial for finance. Future research should explore specialized pre-training objectives that incorporate market microstructure theory and regulatory constraints. The development of modular architectures that combine pre-trained components with task-specific adaptors could enable more efficient transfer learning while maintaining model interpretability.
Multi-agent architectures for portfolio optimization and market making demonstrate strong performance~\cite{li2024masa,zhang2024optimizing} but face challenges in convergence guarantees and Nash equilibrium stability. Research into theoretical frameworks for multi-agent learning under market non-stationarity could lead to more robust trading systems. The integration of market impact models into agent training frameworks, building upon work like~\cite{ijcai2024p664}, remains crucial for bridging the gap between simulation and real-world deployment.

\subsection{Methodological Advancements}
Several methodological challenges require attention from the research community. The treatment of temporal dependencies in financial data remains suboptimal, with current approaches~\cite{wang2023timemixer,Yuqietal-2023-PatchTST} struggling to capture long-range dependencies while maintaining computational efficiency. Future work should explore adaptive attention mechanisms that dynamically adjust their receptive field based on market conditions.
The integration of classical financial theory with deep learning frameworks presents another promising direction. While works like~\cite{xu2024garch} incorporate GARCH models and factor analysis, a more systematic approach to theory-guided neural network design could improve both performance and interpretability. This includes developing loss functions that explicitly encode financial principles and constraints.
Risk modeling in deep learning frameworks requires particular attention. Current approaches~\cite{hui2024mitigating,yamagata2024risk} often rely on simple volatility estimates or Sharpe ratios, failing to capture tail risks and regime changes adequately. Research into neural architectures specifically designed for extreme value theory and systemic risk modeling could significantly improve portfolio management and risk assessment.

\subsection{Multimodal Foundation Models}
The emergence of multimodal foundation models presents exciting opportunities for financial applications. While current work~\cite{li2024alphafin,zhang2024ai} demonstrates promise in combining textual and numerical data, future research should explore more sophisticated fusion mechanisms. This includes developing specialized architectures for processing earnings calls, satellite imagery, and alternative data sources simultaneously.
The adaptation of foundation models for real-time market analysis remains challenging. Research into streaming architectures that can efficiently update their knowledge base and adapt to new market conditions could enable more responsive trading systems. Additionally, exploring techniques for maintaining temporal coherence across different data modalities could improve prediction accuracy during market regime changes.

\subsection{Hardware-Accelerated Trading}
The advancement of hardware-accelerated trading systems presents several research opportunities. Current approaches~\cite{han2023efficient,qin2024earnhft} demonstrate promising potential for low-latency trading. Research into specialized neural network architectures optimized for FPGA implementation could further reduce latency while maintaining model sophistication.
The development of hardware-aware training algorithms represents another promising direction. Future work should explore techniques for jointly optimizing model architecture and hardware implementation, potentially leading to more efficient high-frequency trading systems. This includes developing specialized attention mechanisms and network pruning techniques that consider hardware constraints during training.
The integration of hardware acceleration with market making systems presents unique challenges. Research into architectures that can efficiently process order book data and execute trades with microsecond latency could significantly improve market liquidity. This includes developing specialized circuits for order matching and risk calculation that maintain accuracy while minimizing latency.

\subsection{Research-Industry Collaborations}
A notable limitation in current research is the absence of real-world deployment studies and industry validation. While works like~\cite{han2023efficient} report production latency metrics, and~\cite{qin2024earnhft,zong2024macrohft} demonstrate strong backtesting performance, none of the surveyed papers provide comprehensive evidence of successful industrial deployment. This gap manifests in several critical aspects:
First, most research evaluations rely solely on historical data backtesting, which fails to capture crucial real-world challenges such as market impact, execution slippage, and microstructure effects. The lack of live trading results or paper trading validation raises questions about the practical viability of proposed methods.
Second, computational requirements and system integration receive limited attention. While papers like~\cite{han2023efficient} discuss latency constraints, few address critical production concerns such as system reliability, fault tolerance, and integration with existing trading infrastructure. The absence of deployment case studies or performance analysis under real market conditions represents a significant limitation in current research.
Third, regulatory compliance and risk management frameworks remain largely theoretical. Despite works like~\cite{li2024finreport} addressing interpretability, none of the surveyed papers demonstrate compliance with specific regulatory requirements or integration with institutional risk management systems. This gap between academic innovation and regulatory reality hinders industrial adoption.
Future research would benefit significantly from collaborations between academic institutions and financial firms to validate proposed methods under real market conditions. Studies documenting deployment challenges, system architecture decisions, and practical performance metrics would provide valuable insights for both researchers and practitioners. Additionally, research into robust evaluation frameworks that better approximate real-world trading conditions could help bridge the gap between academic research and industrial applications.

\subsection{Practical Considerations}
Several critical implementation challenges require focused research attention. While synthetic data generation techniques~\cite{xia2024market} help address data limitations, research is needed on frameworks that can rigorously validate model robustness under real market conditions. This includes developing standardized stress testing methodologies that simulate market crises, liquidity shocks, and extreme events.
System resilience and failsafe mechanisms represent another crucial area. Research into graceful degradation strategies, automated circuit breakers, and robust fallback mechanisms could improve the safety of automated trading systems. This includes developing methods to automatically detect and respond to anomalous market conditions or model behavior.
The challenge of continuous learning and model updating in production environments remains largely unexplored. Future research should investigate techniques for safely updating models in live trading systems, including methods for gradual deployment, A/B testing, and performance monitoring. This includes developing frameworks for detecting model degradation and safely reverting changes if necessary.
Cross-jurisdictional compliance presents another significant challenge, particularly for global trading operations. Research into automated compliance verification and real-time regulatory reporting could help bridge the gap between academic innovation and practical deployment. This includes developing standardized interfaces between trading systems and regulatory monitoring frameworks.
These practical considerations highlight the need for research that explicitly addresses production deployment challenges while maintaining the theoretical rigor expected in academic work. Success in these areas could accelerate the adoption of AI innovations in real-world financial systems.
\section{Conclusion}
This survey has systematically analyzed recent advancements in artificial intelligence for financial applications, revealing significant progress across predictive modeling, decision-making, and knowledge retrieval tasks. The examined innovations span architectural developments in foundation models, specialized network designs for temporal and relational learning, and practical frameworks for production deployment. While these advances demonstrate substantial improvements in model performance and capability, several crucial challenges remain unresolved.

The theoretical foundations for financial AI systems still require strengthening, particularly in establishing convergence guarantees for multi-agent trading systems and developing robust frameworks for handling market non-stationarity. The integration of classical financial theory with deep learning architectures presents another critical area for development, as current approaches often fail to fully incorporate market microstructure effects and regulatory constraints.

Production deployment remains a significant challenge, with few studies demonstrating successful real-world implementation or comprehensive compliance with regulatory requirements. The gap between academic innovation and industrial application suggests the need for closer collaboration between researchers and practitioners, particularly in developing standardized evaluation frameworks that better approximate real-world trading conditions.

Future research directions should focus on developing more robust theoretical frameworks for financial AI, improving the integration of domain knowledge in model architecture design, and addressing the practical challenges of system deployment and regulatory compliance. The advancement of hardware-accelerated trading systems and the development of multimodal foundation models present particularly promising opportunities for innovation. Success in these areas could significantly advance the field while improving the reliability and effectiveness of AI-driven financial systems.


\bibliographystyle{named}
\bibliography{intro,portfolio,trend,price-forecast,quant,ranking-based,dataset-dataaug,ts-models}

\end{document}